\documentclass[showpacs,preprint,amsmath,amssymb]{revtex4}
\usepackage{graphicx}
\usepackage{dcolumn}
\usepackage{bm}
\usepackage{color}

\begin{document}


\title{Exact evaluation of the cutting path length in a percolation model on a hierarchical network}
\author{R. F. S. Andrade$^{1}$, H. J. Herrmann$^{2,3}$}
\affiliation{$^1$Instituto de F\'{i}sica,
Universidade Federal da Bahia,
40210-210, Salvador, Brazil.\\
$^{2}$Computational Physics, IfB, ETH-H\"{o}nggerberg, Schafmattstr. 6,
8093, Z\"{u}rich, Switzerland.
\\$^{3}$Departamento de F\'{i}sica, Universidade Federal do Cear\'{a},
Campus do Pici, 60455-760, Fortaleza, Brazil.}

\date{\today}

\begin{abstract}
This work presents an approach to evaluate the exact value of the fractal
dimension of the cutting path $d^{CP}_f$ on hierarchical structures with
finite order of ramification. This represents the first renormalization
group treatment of the universality class of watersheds. By making use of
the self-similar property, we show that $d^{CP}_f$ depends only on the
average cutting path (CP) of the first generation of the structure. For
the simplest Wheastone hierarchical lattice (WHL), we present a
mathematical proof. For a larger WHL structure, the exact value of
$d^{CP}_f$ is derived based on an computer algorithm that identifies the
length of all possible CP's of the first generation.
\end{abstract}
\pacs{05.10.Cc, 64.60.ah, 89.75.Da.}

\maketitle

\section{Introduction}

The evaluation of the watershed of a rough landscape is of fundamental
interest, impacting many different geographic aspects, starting from the
very definition of national and regional limits between countries, to
evaluating the destination of water, pollutants, and agricultural
fertilizers washed out by rain. The challenge of presenting a general
mathematical solution for the watershed given a general landscape is still
open. Recent extensive numerical work provided some quantitative
information characterizing some watershed features, as the fractal
dimension $d^{WS}_f$.

It was observed that the value $d^{WS}_f\simeq1.21...$, which is obtained
for an uncorrelated random landscape \cite{Fehr2009,Fehr2011}, has also
been reported in the evaluation of the fractal dimension of other models,
like the set of cutting bonds below and the set of bridge bonds above the
threshold, respectively, of a percolation problem on a two dimensional
lattice \cite{Schrenk2012a}. The same fractal dimension has also been
obtained for the optimum path crack \cite{Soares2009}, the surface of the
clusters in explosive percolation \cite{Nuno2010}, the random fuse model
in strong disorder \cite{Moreira2012}, the shortest path in loopless
percolation \cite{Porto1997} and the random polymer in high disorder
\cite{Cieplak1994}. As far as we know, the quoted (or related) problems
have never been subject of an exact analysis, although it has been shown
that they fulfill the Schramm-Loewner evolution (SLE) properties
\cite{Daryaei2012}. Therefore, the identification of a universal physical
mechanism justifying the emergence of this particular $d_f$ value for
these problems is an open challenge.

In this work, we present an exact evaluation of the fractal dimension of
the cutting path length on two distinct Wheatstone hierarchical lattices
(WHL) (see Fig. 1). Hierarchical lattices have played an important role in
the study of critical phenomena, since the results they produce can be
regarded as approximations to models on Euclidian lattices within the
framework of the Migdal-Kadanoff real space renormalization group
\cite{Migdal1975a,Migdal1975b,Kadanoff1976,Berker1979,Bleher1979,Kauffman1981}.
The solution we provide follows similar steps as used in the analysis of
physical models on hierarchical structures with finite ramification order.
We make use of the exact scale invariance of the geometrical construction
to analyze successive generations $g$ of the model, requiring that the
same expressions be valid for $g$ and $g+1$ provided the pertinent
quantities are re-scaled in the way dictated by the geometrical
construction.

The WHL hierarchical structure is constructed by successively replacing a
simple line segment between two root points (the $g=0$ generation)  by a
more complex structure consisting of a set of $p$ parallel branches, which
contains $b$ inner connections and $b-1$ sites. The inner sites in each
neighboring parallel branch are also connected by a bond. As a result of
this procedure, the lattice at the $g+1$ generation can also be obtained
by substituting each of the bonds of the $g=1$ generation by a $g$
generation lattice. This way of building up the successive generations of
the lattice will be adopted in the derivation of our main result. The
resulting self-similar graph has a fractal dimension
\cite{Tsallis1996,Almeida2005}

\begin{equation}\label{eq0}
    d^W_f=\log(bp+(b-1)(p-1))/ \log b.
\end{equation}

\noindent In the current work, we consider $b=p=2$ and $b=p=3$ (see Figs.
1 and 2). For these cases, Eq.(\ref{eq0}) leads to $d^W_f\simeq2.322...$
and $d^W_f\simeq2.335...$ correspondingly. $d^W_f$ converges to 2 in a
logarithmic way when $b=p \rightarrow \infty$. The maximal number of bonds
$B$, sites $N$, and the shortest distance between root sites $D$ depend on
$b$ and $g$. When $b=2$ ($b=3$), they are given by $B_g=5^g(13^g)$,
$N_g=(5^g+3)/2((13^g+3)/2)$, and $D_g=2^g(3^g)$. When $b=p$, the WHL's
become self-dual hierarchical structures, in the sense that the dual of
the basic unit is topologically identical to the original structure. This
property has proven to be of relevance in the analysis of spin models. For
instance, it ensures that the WHL critical temperatures coincide with
those of the corresponding models on square lattices, although the
critical exponents are different\cite{Tsallis1996}.

\section{The cutting bond path}

\begin{figure}
\includegraphics*[width=8cm, angle=-90]{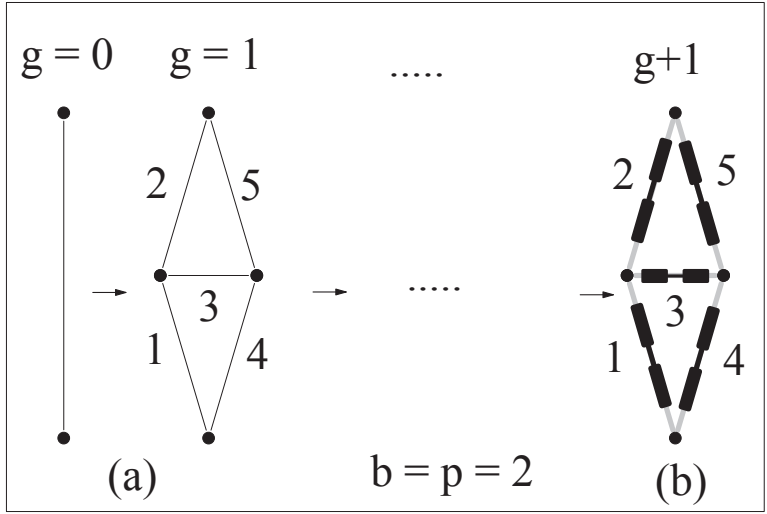}
\caption{(a) First steps of construction of the $b=p=2$ WHL, corresponding
to $g=0$ and $g=1$. (b) Schematic construction of the $g+1$ lattice based
on 5 $g$ lattices, indicated by stylized dumbbells. Single bonds and
dumbbells labeled from 1 through 5 are used in the proof of the main
result in Section III.}\label{fig1}
\end{figure}

\begin{figure}
\includegraphics*[width=6cm, angle=0]{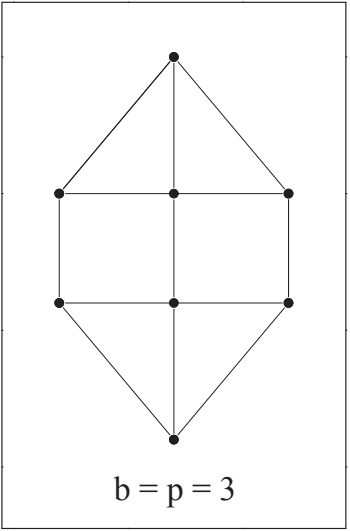}
\caption{First generation of the $b=p=3$ WHL.}\label{fig2}
\end{figure}

The concept of cutting bond path (CP) in a bond percolation model is
defined for a dynamical process where existing bonds are randomly chosen
and removed from the lattice. In this process, the order at which the
bonds are removed is important, which turns the configuration space to
grow as $B!$ instead of the usual dependence $\sim exp(B)$.

For the sake of definitiveness, let us start with a completely filled
square lattice with $B=2L(L-1)$ bonds and $N=L \times L$. This
configuration clearly allows for a connection between any site in the
lowest row to any site in the upper most row.

Bonds are successively eliminated until the choice falls to a bond, which
if removed would disconnect the upper and lower rows. This bond is labeled
as cutting bond (in that particular removal sequence) and will be kept in
the lattice, i.e., not removed, being the first element of the set of
bonds in the CP. The removal process proceeds further, and a bond is only
not removed whenever its elimination would lead to the disruption of the
connection between the two boundary rows. When such an event occurs, the
number of elements in the CP set is increased by one. The process is
repeated until no other bonds but those labeled as cutting bonds remain in
the lattice. The process that identifies the elements in the CP set
ensures that they form a continuous line consisting of $\ell(i)$ bonds
linking one site in the lowest row to one site in the upper most row. The
index $i$ identifies the random sequence at which the bonds should be
successively eliminated. We denote by $S$ the set of all distinct CPs.
Note that a CP is dependent of the order at which the cutting bonds are
assigned to it, so that $S$ contains a large number of elements that are
formed by the same set of bonds. The average cutting path length is
obviously

\begin{eqnarray}\label{eq1}
 \langle \ell \rangle &=& \frac{1}{\Omega(B)}\sum_{i=1}^{\Omega(B)}\ell(i)\\
\nonumber  &=& \frac{1}{\Omega(B)}\sum_{i\in S}\ell(i)
\end{eqnarray}

\noindent where $\Omega(B)=B!$ counts the number of different sequences
according to which the $B$ bonds are considered for elimination from the
lattice. The fact that the value of $\Omega(B)$ increases faster than
exponential is the essential difficulty to be overcome in an exact
calculation of $\langle \ell \rangle$.

\section{The WHL cutting path}

Let us consider the task of evaluating $\langle \ell \rangle$ for the WHL.
For such structure, we have to start by considering all paths that connect
one root site to the other. The random choice of the bonds to be
eliminated follows the same prescription as for the square lattice, a
selected bond becoming a cutting bond if its removal from the structure
causes the connection between the two root sites to break.

The difficulty of this otherwise very hard task can be managed if we take
advantage from the hierarchical structure of WHL and from the fact that
its sites have a finite ramification order. The fact that, even in the
limit $g\rightarrow \infty$, the WHL structure can be decoupled into more
than one infinite component by removing a finite number of sites makes it
possible to recursively evaluate $\langle \ell_g \rangle$ for any
generation $g$. Let us consider the simpler case $b=p=2$ and proceed by
mathematical induction over $g$ to prove that $\langle \ell_{g} \rangle =
\left (\frac{34}{15}\right )^g$.

\subsection{$b=p=2$ and $g=1$}

Let us indicate by $S_g$ the set of all distinct CPs in the WHL at a given
value of $g$. When $g=1$, the number of bonds is $B_1=5$, so that there
are $\Omega(5)=5!=120$ sequences according to which the individual bonds
can be eliminated. We start the computation of the CP lengths by noting
that the 120 sequences can be cast into two distinct sets (${t}_1$ and
${a}_1$), according to whether the \emph{transverse} bond is the first one
to be eliminated or one of the four \emph{adjacent} bonds to one root site
is chosen first.

The set ${t}_1$ is such that, for all of its 24 different sequences, we
obtain $\ell(i)=\ell_{g=1,t}=\langle  \ell_{1,t}\rangle =2$. The set
${a}_1$ comprises 96 sequences but, due to the up-down and right-left
symmetries of the WHL structure, it is necessary to compute only the
subset of sequences starting with the elimination of the bond 1, according
to the labeling indicated in Fig. 1. We let the partial average for the
first elimination of an adjacent bond $\langle \ell_1(a) \rangle$ be
written as:

\begin{equation} \label{eq2}
\langle \ell_{1,a} \rangle=\frac{1}{4}\left ( \ell_1(1,2) + \ell_1(1,3) +
\ell_1(1,4) + \ell_1(1,5) \right ) ,
\end{equation}

\noindent where the subscript 1 denotes the value of $g$ and $\ell_1(1,j)$
indicates the average CP length when the $j$-th bond is second one to be
eliminated. When the bond 2 is eliminated in second place, the CP is
formed by the bonds 4 and 5, irrespective of the order of the elimination
of the three remaining bonds. The same happens when the bond 3 is
eliminated in second place, so that $\ell_1(1,2) = \ell_1(1,3) = 2$ for
all 6 sequences that enter in the formation of each partial average.

If the bond 4 is chosen to be the second one to be removed, the connection
between the root sites would be broken at that step, so that the bond 4 is
the first cutting bond for such sequences. Next, it is necessary to
analyze in detail the third elimination. If either of the bonds 2 or 3 is
chosen, then the CP length is 2 irrespective of the order of the remaining
eliminations. If the bond 5 is eliminated in third place, the CP length is
3, so that $\ell_1(1,4) = (3 + 2\times 2)/3$. Finally, if the bond 5 is
eliminated in second place, the CP length is 3 for all 6 different
sequences for eliminating bonds 2, 3 and 4.

This leads to

\begin{equation} \label{eq3}
\langle \ell_{1,a} \rangle = 7/3
\end{equation}

\noindent and, as a consequence,

\begin{equation}
\label{eq4}
\langle \ell_1\rangle =\frac{1}{5}\left ( \langle \ell_{1,t}\rangle + 4
\langle \ell_{1,a}\rangle \right )=\frac{34}{15}.
\end{equation}

\noindent Since the $g=0$ WHL consists of the single direct connection
between the root sites, it is obvious that $\langle \ell_0 \rangle = 1$
and, as consequence, our statement is verified for $g=1$.

Before we proceed further with our proof, let us remind that the $g=2$ WHL
generation contains $B_2=25$ bonds, so that the number of different
sequences over which the average of $\ell(i)$ has to be performed is
$\Omega(25)=25!>10^{25}$. This provides a flavor for the difficulties one
faces to evaluate $\langle \ell_g \rangle$ if we could not use the
arguments based on the geometric symmetry and finite ramification order.

\subsection{$b=p=2$ and $g>1$}

The second part of the proof depends essentially on the fact that the five
$g$ lattices forming the $g+1$ lattice are separated by cutting sites, so
that each $g$ unit is independent from the others. Figure 1b illustrates,
in a schematic way, how the $g+1$ lattice is assembled by connecting five
$g$ lattices, which are indicated by dumbbells. Next we note that, once a
cutting bond emerges in one of the $g$ lattices, then the two root sites
of that $g$ lattice will be connected by a part of the cutting path in the
$g+1$ lattice. The opposite is also valid, in the sense that either the
two root sites of a $g$ lattice are connected by the cutting path or there
will be no cutting bond in that $g$ lattice. As long as the first cutting
bond does not appear in the sequence of bond eliminations, the order at
which the bonds are removed is only important within each $g$ lattice, and
does not depend on the relative order of elimination among the five
lattices. The order at which bonds are eliminated in other units only
becomes relevant when two (or three) units contain at least one cutting
bond. However, even in this situation, the form of the cutting path within
one $g$ lattice does not depend on the order of the bond elimination in
other $g$ lattices with cutting bonds.

Once these geometrical features have been clarified, we proceed with some
mathematical rigor to complete the proof by assuming that $\langle \ell_g
\rangle=\left (\frac{34}{15}\right )^g$. Our purpose is to show that
$\langle \ell_{g+1} \rangle=\left (\frac{34}{15}\right )^{g+1}$, which
requires that we adapt the counting process used in the previous
Sub-section.

We start by noting that $\Omega(B_{g+1})$, the number of different
sequences in generation $g+1$, increases by a factor
\begin{equation}
    \frac{(5^{g+1})!}{(5^{g})!}=\prod_{k=5^{g}+1}^{k=5^{g+1}}k \nonumber
\end{equation}
\noindent with respect to $\Omega(B_{g})$. This very large factor is due
to the fact that the random bond selection process does not require that
each of the five generation $g$ lattices that are put together to build
one generation $g+1$ lattice be sequentially emptied. If this was the
case, this factor would simply be $5!$. However, the identification of
individual removal sequences is not necessary. In fact, it is sufficient
to characterize the five sets $E_{g+1}(k),$ $k=1,2,3,4,5$, respectively
containing the removal sequences in which the root sites of the $k$-th $g$
lattice become disconnected in first place. Given the fact that all $g$
lattices are equivalent, it follows that $\mathcal{C}(E_{g+1}(k))$, the
number of elements (or cardinality) of each set $E_{g+1}(k)$ is the same
for all values of $k$, what necessarily leads to
$\mathcal{C}(E_{g+1}(k))=\Omega(B_{g+1})/5$.

Now, it follows that there are again $5!$ different sequences along which
the root sites of the five $g$ lattices become disconnected. They are
exactly those identified in the previous Subsection for removing the five
bonds for the $g=1$ generation. Therefore, it is possible to cast all
sequences into a finer classification than that given above, by assigning
the order at which the root sites of each $g$ lattice become disconnected.
We can use the same symmetry arguments used before to identify that there
are, in fact, just two independent five step sequences: the 24 $t$
sequences, at which the transverse $g$ lattice is the first one to be
disconnected, and the 96 $a$ sequences, in which a $g$ lattice sharing one
of its root sites with the $g+1$ lattice is the first one to be
disconnected. Note that the total number of removal sequences contained in
the 24 $t$ sequences is exactly $\Omega(B_{g+1})/5$. If we let $(\langle
\ell_{g+1,t}\rangle$ and $\langle \ell_{g+1,a}\rangle$ be the average
lengths associated with the $t$ and $a$ sequences in the $g+1$ generation,
it is possible to write

\begin{equation}
\label{eq5}
\langle \ell_{g+1}\rangle =\frac{1}{5}\left ( \langle \ell_{g+1,t}\rangle + 4
\langle \ell_{g+1,a}\rangle \right ).
\end{equation}

Let us evaluate $\langle \ell_{g+1,t}\rangle$. Since the transverse $g$
lattice has been disconnected in first place, it turns out that the $g+1$
CP is necessarily formed by bonds placed either in the 1 and 2 $g$
lattices or in the 4 and 5 $g$ lattices. Let us identify the sets of such
sequences by $S_{g+1}(1,2)\subset S_{g+1}$ and  $S_{g+1}(4,5)\subset
S_{g+1}$. Since these are equivalent, we may concentrate, without loss of
generality, on the CP's formed by sequences in $S_{g+1}(1,2)$. It is then
clear that $\mathcal{C}(S_{g+1}(1,2))=\Omega(B_{g+1})/10$, leading to

\begin{eqnarray}
\label{eq6}
\langle \ell_{g+1,t}\rangle &=&\frac{5}{\Omega(B_{g+1})}\sum_{i\in E_{g+1}(3)} \ell_{g+1}(i) \\
\nonumber  &=&\frac{10}{\Omega(B_{g+1})}\sum_{i\in S_{g+1}(1,2)} \ell_{g+1}(i).
\end{eqnarray}

Next we make use of the fact that each CP in the set $S_{g+1}(1,2)$ can be
separated in two parts, each of them formed, respectively, by bonds in the
1 and 2 $g$ lattices. Since all CP's must go through a root site of both 1
and 2 $g$ lattices, a CP in the set $S_g(1,2)$ can be obtained by the
concatenation of two CP's of the generation $g$. Thus, $\ell_{g+1}(i) =
\ell^1_{g,g+1}(i) + \ell^2_{g,g+1}(i)$, where $\ell^j_{g,g+1}(i)$, with
$j=1$ and $2$, is the number of cutting bonds of the $g+1$ sequence $i$ in
the $j$-th $g$ lattice. Therefore we can write

\begin{eqnarray}
\label{eq7}
\langle \ell_{g+1,t}\rangle &=& \frac{10}{\Omega(B_{g+1})}\sum_{i\in S_{g+1}(1,2)} \ell^1_{g,g+1}(i) \\
\nonumber    &+& \frac{10}{\Omega(B_{g+1})}\sum_{i\in S_{g+1}(1,2)} \ell^2_{g,g+1}(i).
\end{eqnarray}

To evaluate the first sum in Eq. (\ref{eq7}), we take into account the
fact that it is possible to identify and separate the independent
contributions depending on the removal sequence inside the $g$ lattices 1
only. Given the fact that the two sums are equivalent, the same procedure
applies also for the second sum. Thus, select one removal sequence $i^*\in
S_{g+1}(1,2)$ and follow the steps at which the specific bonds appear in
the 1 $g$ lattice. Repeat this procedure for each value of $i\leq
\Omega(B_{g+1})/10$. During this process, it is possible to find a large
number $\rho_{g+1}(1,2;i^*)$ of different CPs in the set $S_{g+1}(1,2)$,
for which the subset of bonds in the $j=1$ $g$ lattice is the same as that
in the $i^*$ path. Since this is true for any choice of the path $i^*$, we
can simplify the notation and just write to
$\rho_{g+1}(1,2;i^*)=\rho_{g+1}(1,2)$.

Finally, as the 1 and 2 $g$ lattices are equivalent, we must have
$\rho_{g+1}(1,2)=\rho_{g+1}(2,1)=\Omega(B_{g+1})/10\Omega(B_{g})$.
Therefore, Eq. (\ref{eq7}) can be written as

\begin{eqnarray}
\label{eq8}
  \langle \ell_{g+1,t}\rangle &=& \frac{1}{\Omega(B_{g})\rho_{g+1}(1,2)}\sum_{i\in S_{g+1}(1,2)} \ell^1_{g,g+1}(i) \\
\nonumber    &+& \frac{1}{\Omega(B_{g})\rho_{g+1}(2,1)}\sum_{i\in S_{g+1}(1,2)} \ell^2_{g,g+1}(i).
\end{eqnarray}

\noindent Since the term in the first sum depends only on the bonds in the
1 $g$ lattice, $\sum_{i\in S_{g+1}(1,2)}
\ell^1_{g,g+1}(i)=\rho_{g+1}(1,2)\sum_{i'\in S_g} \ell^1_{g,g+1}(i')$,
where we have used a new variable $i'$ in the sum on the r.h.s to stress
the fact that now the sum is performed over the CP's of the generation
$g$. This property can be immediately used to simplify the second sum, in
such way that

\begin{eqnarray}
\label{eq9}
  \langle \ell_{g+1,t}\rangle &=& \frac{1}{\Omega(B_{g})}\sum_{i'\in S_g} \ell^1_{g,g+1}(i')
  + \frac{1}{\Omega(B_{g})}\sum_{i'\in S_g} \ell^2_{g,g+1}(i')\\
\nonumber    &=& 2 \frac{1}{\Omega(B_{g})}\sum_{i'\in S_g} \ell^2_{g,g+1}(i')\\
\nonumber    &=& 2\langle \ell_{g}\rangle.
\end{eqnarray}

Similar arguments can be used to evaluate $\langle \ell_{g+1,a}\rangle$.
As we have shown in the former Subsection, the CP's resulting from those
removal sequences where the first pair of root sites of a $g$ lattice to
be disconnected have a common root site with the $g+1$ lattice may have
bonds in two or three different $g$ lattices. The proportion of paths in
each situation is exactly the same as that obtained in Eq. (\ref{eq2}), so
that we obtain

\begin{equation} \label{eq10}
\langle \ell_{g+1,a} \rangle = 7/3\langle \ell_{g}\rangle
\end{equation}

\noindent and, as a consequence,

\begin{equation}
\label{eq11}
\langle \ell_{g+1}\rangle =\frac{34}{15}\langle \ell_{g}\rangle.
=\left (\frac{34}{15} \right )^{g+1}.\end{equation}

The above arguments can be used whenever we consider exact self-similar
structures with finite order of ramification. In such cases, the presence
of cutting points in the $g+1$ lattice that are equivalent to the root
sites of the $g$ lattice constitutes the key property that allows for
similar proofs. Therefore, for this class of structures, the value of
$\langle \ell_{g}\rangle$ depends only on the value of $\langle
\ell_{g=1}\rangle$.

\subsection{Fractal dimension}

Since the typical length scale in the WHL is the shortest distance between
the two root sites, it follows that the CP fractal dimension is given by

\begin{equation}
\label{eq12}
d^{CP}_f =\frac{\log(\langle \ell_{g}\rangle)}{\log(D_g)}=\frac{\log(34/15)^g}{\log(2)^g}\simeq 1.1805
\end{equation}

\noindent for the $b=2$ WHL. This value is not far to the one found for
$d^{CP}_f$ in the square lattice.

The evaluation of $d^{CP}_f$ for the $b=3$ WHL is much more difficult,
although we actually need only to evaluate $\langle \ell_{g=1}\rangle$.
The $g=1$ lattice has 13 bonds, hence $13!\sim 6\times 10^{9}$ cutting
paths should be identified. The evaluation of the $\langle
\ell_{g=1}\rangle$ was parallelized and distributed into 25 Xeon CPU
cores, requiring $\sim$ 5 days of CPU time for this task. We obtained the
result

\begin{equation}
\label{eq12}
d^{CP}_f =\frac{\log(\langle \ell_{g}\rangle)}{\log(D_g)}=\frac{\log(35318809/9266400)^g}{\log(3)^g}\simeq 1.21791,
\end{equation}

\noindent which is very close to the accepted value for $d^{CP}_f$ in the
square lattice, which is $1.2168 \pm 0.0005$ \cite{Fehr2012}.

Despite the hierarchical structure of the WHL family and the fact that the
fractal dimension is slightly larger than two, it is interesting to note
that $d^{CP}_f$ for two WHL's are comparable to those found on the square
lattice. It should also be remarked that the next case $b=p=4$ has 25
connections when $g=1$. Note that $25!>10^{25}$, and therefore it seems
that the exact evaluation of $d^{CP}_f$ for larger values of $b$ and $p$
is not feasible with current day computing facilities, at least within the
framework developed herein.

\section{Conclusions}

We have presented the first renormalization group treatment for the
fractal dimension of the cutting bond length, which is in the same
universality class as the watershed, the optimum path crack and the
shortest path in loopless percolation. This calculation is in fact
identical to the exact solution on a specific type of hierarchical
lattice. Our result is particularly interesting because, as opposed to
usual problems in statistical physics, the phase space here grows like
$N!$ instead of exponentially because of the history dependence of the
configurations. It would be interesting to apply in the future our
technique to other problems involving ranked surfaces like, for instance,
the sharing of reservoirs \cite{Schrenk2012b}.

{\bf Acknowledgement}: The authors acknowledge support from the European
Research Council (ERC) Advanced Grant 319968-FlowCCS, from the Brazilian
Agencies FAPESB (project PRONEX 0006/2009) and CNPq, and from the
Brazilian National Institute of Science and Technology of Complex Systems
(INCT-SC).

\bibliographystyle{prsty}

\end{document}